\documentclass[12pt]{iopart}
\usepackage{latexsym}

\def\cH{{\cal H}} 
\def\cR{{\cal R}}

\def\Tr{{\rm Tr}}
\def\d={\buildrel \rm def \over =}

\def\ket#1{\mid~\!\!\!{#1}~\!\!\rangle}
\def\bra#1{\langle~\!\!{#1}~\!\!\!\mid}

\begin{document}

\title{\bf Strong twin events in
mixed-state entanglement}
\author{F Herbut\footnote[1]{E-mail:
fedorh@infosky.net}}
\address{Faculty of Physics, University of
Belgrade, POB 368, Belgrade 11001,
Yugoslavia and The Serbian Academy of
Sciences and Arts, Knez Mihajlova 35,
11000 Belgrade}

\date{\today}

\begin{abstract}
Continuing the study of mixed-state
entanglement in terms of
opposite-subsystem observables the
measurement of one of which amounts to
the same as that of the other
(so-called twins), begun in a recent
article, so-called strong twin events,
which imply biorthogonal mixing of
states, are defined and studied. It is
shown that for each mixed state there
exists a Schmidt canonical (super state
vector) expansion in terms of Hermitian
operators, and that it can be the
continuation of the mentioned
biorthogonal mixing due to strong
twins. The case of weak twins and
nonhermitian Schmidt canonical
expansion is also investigated. A
necessary and sufficient condition for
the existence of nontrivial twins for
separable states is derived.
\end{abstract}


\submitted \maketitle

\section{Introduction}

\rm It was argued in a recent article
\cite{HD00} that the study of
entanglement through twin observables,
or shortly {\it twins}, is important
for quantum communication and quantum
information theories because it reveals
very basic properties. Twin observables
are opposite-subsystem observables such
that the (subsystem) measurement of one
of them amounts to a measurement also
of the other. Equivalently put, the
subsystem measurement of a twin gives
rise, on account of entanglement, to an
orthogonal state decomposition of the
state of the opposite subsystem.

When a general, i. e., mixed or pure,
composite-system state (statistical
operator) $\rho_{12}$ is given, twins
$(A_1,A_2)$ are algebraically defined
as Hermitian (opposite subsystem)
operators satisfying $$
A_1\rho_{12}=A_2\rho_{12},\eqno{(1)}$$
where $A_1$ is actually $A_1\otimes
I_2$, $I_2$ being the identity operator
for the second subsystem, etc. It was
shown \cite{HD00} that (1) implies $$
[A_1,\rho_1]=0,\qquad
[A_2,\rho_2]=0,\eqno{(2a,b)}$$
where
$\rho_1\equiv \Tr_2\rho_{12}$ and
$\rho_2$ (defined symmetrically) are
the subsystem states (the reduced
statistical operators). The symbols
$\Tr_i$, $i=1,2$, denote the partial
traces. Further, the so-called {\it
detectable parts} $A'_i$  of the twins,
the restrictions of $A_i$ to the ranges
$\cR(\rho_i)$, $i=1,2$,
 have equal and
necessarily purely discrete spectra
(but with possibly different
multiplicities of the characteristic
values except in the pure-state case).
Further, the characteristic events
(projectors) corresponding to the same
characteristic value are twins.

Let $(P_1,P_2)$ be a pair of nontrivial
twin events (twin projectors) for
$\rho_{12}$. Then we can decompose the
statistical operator: $$
\rho_{12}=P_1\rho_{12}+P_1^{\perp}
\rho_{12},\eqno{(3)}$$
where
$P_1^{\perp}$ is the orthocomplementary
projector of $P_1$. In general, the
terms on the RHS are not even
Hermitian. First, we are going to
investigate the more important case
when (3) is a mixture of states.

\section{Strong twin projectors and
biorthogonal mixtures}

Let $(P_1,P_2)$ be a pair of nontrivial
twin projectors for a composite-system
statistical operator $\rho_{12}$.\\

\noindent {\it Remark.} Evidently,
either both terms on the RHS of (3) are
Hermitian or none of them. They are
{\it Hermitian} if and only if the
projector $P_1$ (or equivalently,
$P_1^{\perp}$) {\it commutes} with
$\rho_{12}$: $$
[P_i,\rho_{12}]=0,\qquad i=1,2,
\eqno{(4)}$$
(any one of the equalities
implies the other), as seen by
adjoining the terms in (3).

Hermiticity of the terms in (3) implies
that they are statistical operators (up
to normalization constants), i. e.,
that (3) is {\it a mixture}. Namely, if
(4) is valid, then idempotency leads to
$P_1\rho_{12}=P_1\rho_{12}P_1$, which
is evidently a positive operator. Since
$$ \Tr P_1\rho_{12}P_1\leq \Tr
\rho_{12}=1, $$
the operator has a
finite trace.\\

\noindent {\it Definition 1.} We call
nontrivial twin events (projectors)
either {\it strong twin events}
(projectors), if they satisfy (4), or
{\it weak} twin events (projectors) if
(4) is not satisfied.\\

A strong twin event $P_1$ implies a
mixture (3) of states that have a
strong property called biorthogonality.
To understand it, we first remind of
(ordinary) orthogonality of states.

If $\rho'$ and $\rho''$ are statistical
operators with $Q'$ and $Q''$ as their
respective range projectors, then one
has the known equivalences: $$ \rho'
\rho'' =0\quad \Leftrightarrow \quad
Q'Q''=0\quad \Leftrightarrow \quad
\cR(\rho')\perp
\cR(\rho''),\eqno{(5)}$$
where the last
relation expresses orthogonality of the
ranges.

Any of the three relations in (5)
defines {\it orthogonality of
states}.\\

\noindent {\it Definition 2.} If $$
\rho_{12}=w\rho_{12}'+
(1-w)\rho_{12}'',\qquad 0<w<1,
\eqno{(6)}$$ is a mixture of states
such that $$ \rho'_i\rho''_i=0,\qquad
i=1,2,\eqno{(7)}$$
where $\rho'_1\equiv
\Tr_2\rho'_{12}$ etc. are the reduced
statistical operators, then we say that
(6) is a {\it biorthogonal mixture}.\\

To prove a close connection between
strong twin events and biorthogonal
mixtures, we need another known general
property of composite-system
statistical operators $\rho_{12}$: $$
\rho_{12}=Q_1 \rho_{12}=
\rho_{12}Q_1=Q_2 \rho_{12}=
\rho_{12}Q_2,\eqno{(8)}$$
where $Q_i$
is the range projector of the
corresponding reduced statistical
operator $\rho_i$, $i=1,2$.\\

\noindent {\it Theorem 1.} If $P_1$ is
a nontrivial twin event, (3) is a {\it
biorthogonal mixture if and only if}
$P_1$ is a {\it strong twin event}.\\

\noindent {\it Proof. Sufficiency.} If
$P_1$ is a strong twin projector and
(6) is obtained by rewriting (3), then
$w\rho'_{12}=P_1\rho'_{12}$ is valid,
and this implies $\rho'_1=P_1\rho'_1$
for the reduced statistical operator,
and, adjoining this, one arrives at
$\rho'_1=\rho'_1P_1$. On the other
hand, one has analogously
$\rho''_{12}=P_1^{\perp}\rho''_{12}$
implying $\rho'_1=P_1^{\perp}\rho'_1$.
Finally,
$$
\rho'_1\rho''_1=(\rho'_1P_1)
(P_1^{\perp}\rho''_1)=0.$$

The symmetrical argument holds for the
second tensor factor.

{\it Necessity.} If (6) is a
biorthogonal mixture, then we define
$P_i\equiv Q'_i$, $i=1,2$, i. e., we
take the range projectors of the
reduced statistical operators of
$\rho'_{12}$ as candidates for our twin
projectors. On account of (8), we can
write (6) as follows:
$$
\rho_{12}=wQ_1'Q_2' \rho_{12}'Q_1'Q_2'+
(1-w)Q_1''Q_2''
 \rho_{12}''Q_1''Q_2''.$$
Since in view of (5) biorthogonality
(7) implies $Q_i'Q_i''=0$, $i=1,2$, it
is now obvious that $P_1$ and $P_2$,
multiplying from the left $\rho_{12}$,
give one and the same operator, i. e.,
that they are twins, and it is also
obvious that they both give the same
irrespectively if they multiply $
\rho_{12}$ from the left or from the
right, i. e., that they are strong twin
projectors.

\hfill $\Box$

In view of (5), it is clear that
biorthogonal decomposition of a
statistical operator can be, in
principle, {\it continued}: If, e. g.,
$\rho'_{12}$ in a biorthogonal
decomposition (6) is, in its turn,
decomposed into biorthogonal
statistical operators and replaced in
(6), then any two of the new terms are
biorthogonal etc.

An extreme case of a biorthogonal
mixture is a {\it separable} one: $$
\rho_{12}=\sum_kw_k\Big(\rho_1^{(k)}\otimes
\rho_2^{(k)}\Big),\eqno{(9)}$$ where $$
\forall k:\quad w_k>0,\enskip
\rho_i^{(k)}>0,\enskip
\Tr\rho_i^{(k)}=1,\enskip i=1,2;\quad
\sum_kw_k=1$$ ("$\rho >0$" denotes
positivity of the operator). This
decomposition cannot, of course, always
be carried out, but examples are well
known. For instance, if one performs
ideal measurement of the z-component of
spin of the first particle in a singlet
two-particle state, one ends up with $$
\rho_{12}\equiv
(1/2)\Big(\ket{z+}_1\bra{z+}_1 \otimes
\ket{z-}_2\bra{z-}_2\enskip +\enskip
\ket{z-}_1\bra{z-}_1 \otimes
\ket{z+}_2\bra{z+}_2\Big).$$
This is
obviously a biorthogonal separable
mixture.

One wonders if, at the price of
relaxing the requirement of
statistical-operator terms as slightly
as possible, there could exist a {\it
general} decomposition into {\it
uncorrelated} terms (like in (9)).

To find an affirmative answer, we take
resort to the known case of general
(entangled or disentangled)
composite-system {\it state vectors}
and their Schmidt canonical expansions.
Let us sum up the relevant information
on this \cite{FV76}.

{\it The Schmidt canonical expansion}
(also called Schmidt biorthogonal
expansion) of an arbitrary pure state
vector $\ket{\Phi}_{12}$ of a composite
system is expressed in terms of {\it
its canonical entities}. They are the
following:

(i) {\it The reduced statistical
operators} (subsystem states) $\rho_1$
$\Big(\equiv
\Tr_2\ket{\Phi}_{12}\bra{\Phi}_{12}\Big)$
and $\rho_2$ (defined symmetrically)
are well known.

(ii) The spectral forms of the reduced
statistical operators are $$
\rho_1=\sum_ir_i\ket{i}_1\bra{i}_1,
\quad
\rho_2=\sum_ir_i\ket{i}_2\bra{i}_2,
\quad \forall i:\enskip r_i>0.
\eqno{(10a,b)}$$
(Note that the
positive spectra -multiplicities
included - are always equal.)

(iii) Finally, the mentioned expansion
utilizes the (antiunitary ) {\it
correlation operator} $U_a$, which maps
the range $\cR(\rho_1)$ onto the range
$\cR(\rho_2)$. (Note that they are
always equally dimensional in the pure
state case). The correlation operator
is determined by $\ket{\Phi}_{12}$,
and, in turn, in conjunction with
$\rho_1$, it determines
$\ket{\Phi}_{12}$.

The {\it Schmidt canonical expansion}
reads: $$
\ket{\Phi}_{12}=\sum_ir_i^{1/2}
\ket{i}_1\otimes
\Big(U_a\ket{i}_1\Big)_2. \eqno{(11)}$$
The normalized characteristic vectors
$\ket{i}_2$ in (10b) may (and need not)
be chosen to be equal to
$\Big(U_a\ket{i}_1\Big)_2$.

\section{Hermitian Schmidt canonical
expansion of statistical operators}

It is well known that linear
Hilbert-Schmidt operators $A$, i. e.,
those with a finite Hilbert-Schmidt
norm $\Big(\Tr
A^{\dagger}A\Big)^{1/2}$, form a
Hilbert space in their turn. Writing
the operator $A$ as a (Hilbert-Schmidt)
supervector $\ket{A}$, the scalar
product is
$$ \bra{A}\ket{B}\equiv \Tr
A^{\dagger}B.$$

Since for every statistical operator
$\rho$, one has $\Tr\rho^2\leq 1$, it
is a  Hilbert-Schmidt operator.
Therefore, {\it every statistical
operator has a Schmidt canonical
expansion}.

The trouble is that the operators that
take the place of the statistical
operators $\rho_i^{(k)},\enskip i=1,2$
in (9) are, in general, linear
operators. This might be a too wide
generalization. One wonders if one
could be confined to Hermitian
operators.

When we view the operators as
supervectors, then we must view {\it
adjoining} of operators as {\it an
antiunitary} operator the square of
which is the identity operator, i. e.,
which is an {\it involution}. Hence, we
denote adjoining by $V_1^{(a)}\otimes
V_2^{(a)}$ for a composite system.
Hermitian are the operators that are
{\it invariant} under the action of
this antiunitary involution.

Fortunately, the Schmidt canonical
expansion can always be expressed in
terms of Hermitian operators. We put
this in a more precise and a more
detailed way. But it is simpler to
return to the Hilbert space of state
vectors.\\

\noindent {\it Theorem 2.} Let
$V_1^{(a)}\otimes V_2^{(a)}$ be a given
antiunitary involution acting on
composite-system state vectors. One has
the equivalence:
$$\Big(V_1^{(a)}\otimes
V_2^{(a)}\Big)\ket{\Phi}_{12}=\ket{\Phi}_{12}
\quad \Leftrightarrow \quad
[\rho_i,V_i^{(a)}]=0,\enskip
i=1,2;\quad
V_2^{(a)}U_aV_1^{(a)}=U_a,\eqno{(12)}$$
where $\rho_i,\enskip U_a$ are the
above mentioned canonical entities of
$\ket{\Phi}_{12}$. (Note that in the
last relation we, actually, have the
restriction of $V_1^{(a)}$ to
$\cR(\rho_1)$.)\\

\noindent {\it Proof.} Let
$\ket{\Phi}_{12}$ be invariant under
the action of the antiunitary
involution. Then $$
V_1^{(a)}\rho_1V_1^{(a)}=
V_1^{(a)}\Big(\Tr_2
\ket{\Phi}_{12}\bra{\Phi}_{12}\Big)
V_1^{(a)}=$$ $$ \Tr_2\Big(V_1^{(a)}
\ket{\Phi}_{12}\bra{\Phi}_{12}
V_1^{(a)}\Big)=\Tr_2\Big[V_1^{(a)}
\Big(V_1^{(a)}\otimes V_2^{(a)}\Big)
\ket{\Phi}_{12}\bra{\Phi}_{12}
\Big(V_1^{(a)}\otimes V_2^{(a)}\Big)
V_1^{(a)}\Big]=$$ $$\Tr_2\Big(
V_2^{(a)}
\ket{\Phi}_{12}\bra{\Phi}_{12}
V_2^{(a)}\Big)=\Tr_2
\ket{\Phi}_{12}\bra{\Phi}_{12}=
\rho_1,$$
and symmetrically for
$\rho_2$. One has to note that an
antiunitary involution equals its
inverse and its adjoint. Further, use
has been made of some known basic
properties of partial traces (which are
analogous to the well known ones for
ordinary traces).

Commutation of $\rho_1$ with
$V_1^{(a)}$ allows one to choose the
characteristic basis
$\{\ket{i}_1:\forall i\}$ of the former
spanning its range consisting of
vectors invariant under the action of
$V_1^{(a)}$ (cf \cite{Mes}).

Now, let us take the Schmidt canonical
expansion (11) in terms of an invariant
basis. Then $$ (V_1^{(a)}\otimes
V_2^{(a)})
\ket{\Phi}_{12}=\sum_ir_i^{1/2}
\ket{i}_1\otimes
V_2^{(a)}\Big(U_a\ket{i}_1\Big)_2.$$
Since $\ket{\Phi}_{12}$ was assumed to
be invariant, we have also $$
\ket{\Phi}_{12}=\sum_ir_i^{1/2}
\ket{i}_1\otimes V_2^{(a)}
\Big(U_a\ket{i}_1\Big)_2.$$ The second
tensor factor in each term is uniquely
determined by the LHS and the
corresponding first tensor factor (as a
partial scalar product, cf
\cite{FV76}). Comparison with (11) then
shows that $$ \forall i:\qquad
V_2^{(a)}U_a\ket{i}_1=U_a\ket{i}_1.$$
Since $\ket{i}_1=V_1^{(a)}\ket{i}_1$,
we further have $$
V_2^{(a)}U_aV_1^{(a)}=U_a$$
as claimed.

Conversely, if the main canonical
entities are in the relation to the
antiunitary involutions as stated in
(12), then we can expand
$\ket{\Phi}_{12}$ in a characteristic
basis of $\rho_1$ spanning its range
that is invariant under the antilinear
operator. Then (11) immediately reveals
that, as a consequence,
$\ket{\Phi}_{12}$ is invariant under
$V_1^{(a)}\otimes V_2^{(a)}$. \hfill
$\Box$\\

\noindent {\it Corollary 1.} Every
composite-system statistical operator
$\rho_{12}$ has a {\it Hermitian}
Schmidt canonical expansion.\\

\noindent {\it Proof.} Since every
$\rho_{12}$, being Hermitian, is
invariant under the antiunitary
involution $V_1^{(a)}\otimes
V_2^{(a)}$, Theorem 2 immediately
implies that $\rho_{12}$, upon super
vector normalization, has a Schmidt
canonical expansion in terms of
Hermitian operators.\hfill $\Box$\\

 Returning to a
biorthogonal mixture, one wonders if
one can {\it continue} such a
decomposition by writing each term in a
Hermitian Schmidt canonical expansion
in order to obtain the latter expansion
for the entire statistical operator.
The answer is affirmative on account of
the following result.

Going back to (5), we add a fourth
equivalent property.\\

{\it Proposition 1.} Two statistical
operators $\rho'$ and $\rho''$ are {\it
orthogonal if and only if they are
orthogonal as Hilbert-Schmidt
supervectors}.\\

{\it Proof.} It is obvious that
orthogonality (in the sense of (5))
implies Hilbert-Schmidt orthogonality.
To see the converse implication, we
make use of the fact that every
statistical operator has a purely
discrete spectrum \cite{Sim}, and we
decompose the statistical operators in
terms of characteristic vectors
corresponding to positive
characteristic values: $$
\bra{\rho'}\ket{\rho''}=\Tr\rho'
\rho''=\Tr\sum_kr_k\ket{k}\bra{k}
\sum_j\bar r_j\ket{j}\bra{j}=$$ $$
\sum_k\sum_jr_k\bar r_j
|\bra{j}\ket{k}|^2.$$ Hence, $$
\bra{\rho'}\ket{\rho''}=0\quad
\Rightarrow \quad \rho'\rho''=0$$
(cf
the third relation in (5)).\hfill
$\Box$\\

 If $(A_1,A_2)$ is a pair of {\it twin
observables}, then, as it was stated
(cf also \cite{HD00}), the detectable
parts $A_i',\enskip i=1,2$, have a
common purely discrete spectrum
$\{a_n:\forall n\}$ (with, in general,
different multiplicities), and the
corresponding characteristic projectors
$\{P_i^{(n)}:i=1,2\enskip \forall n\},$
are also pairs of twins.\\

\noindent {\it Definition 3.} If {\it
all} mentioned characteristic projector
pairs $(P_1^{(n)},P_2^{(n)})$ are
strong twin projectors, then
$(A_1,A_2)$ is a pair of {\it strong
twin observables}. If some of the
detectable characteristic twin
projectors are strong and some weak, we
say that we have {\it partially strong}
(or, synonymously, partially weak) twin
observables. If all the mentioned twin
projectors are weak, then we have a
{\it weak} pair of twin observables.\\

Evidently,  a pair $(A_1,A_2)$ of
nontrivial twin observables for
$\rho_{12}$ is a pair of strong ones if
and only if $$ [A_i,\rho_{12}]=0,\quad
i=1,2\eqno{(13)}$$
is valid. This is so
because commutation with all
characteristic projectors is equivalent
to commutation with the Hermitian
operator itself.

Strong twin observables, by means of
their strong characteristic twin
projectors, lead to a generalization of
(3): $$
\rho_{12}=\sum_nP_1^{(n)}\rho_{12}=
\sum_nw_n\rho_{12}^{(n)},\eqno{(14a)}$$
where $$ \forall n:\qquad w_n\equiv \Tr
\rho_{12}P_1^{(n)},\quad
\rho_{12}^{(n)}\equiv (w_n)^{-1}
P_1^{(n)}\rho_{12},\eqno{(14b)}$$ and
any two terms in (14a) are
biorthogonal. (Note that we utilize the
entire characteristic projectors, which
are the orthogonal sums $P_1^{(n)}=
(P_1')^{(n)}\oplus (P_1'')^{(n)}$
parallelling $\cH_1=\cR(\rho_1)\oplus
\cR^{\perp}(\rho_1)$ because
$(P_1')^{(n)}\rho_{12}=P_1^{(n)}\rho_{12}$.)
\\

\noindent {\it Proposition 2.} If $$
\rho_1^{(n)}\equiv
\Tr_2\rho_{12}^{(n)},$$ and
symmetrically for $\rho_2^{(n)}$, are
the reduced statistical operators of
the terms in the biorthogonal mixture
(14a), then $$
P_i^{(n)}\rho_i^{(n)}=\rho_i^{(n)},\quad
i=1,2,\eqno{(15a)}$$ or equivalently,
$$ \cR(\rho_i^{(n)})\subseteq
\cR(P_i^{(n)}),\quad
i=1,2.\eqno{(15b)}$$

{\it Proof.} On account of the
definition of (14a), one has
$P_i^{(n)}\rho_{12}^{(n)}=\rho_{12}
^{(n)}$. Taking the opposite-subsystem
partial trace, one obtains
$P_i^{(n)}\rho_i^{(n)}=\rho_i
^{(n)}\enskip i=1,2$.\hfill $\Box$\\

\noindent {\it Corollary 2.} If the
detectable part $A'_1$ of a twin
observable $A_1$ has a {\it
nondegenerate} characteristic value
$a_n$ corresponding to a strong
characteristic twin projector
$(P_1')^{(n)}=\ket{\psi^{(n)}}_1
\bra{\psi^{(n)}}_1,\quad
\ket{\psi^{(n)}}_1\in \cR(\rho_1)$,
then the term in the biorthogonal
mixture (14a) that corresponds to it
has the form $$
w_n\ket{\psi^{(n)}}_1\bra{\psi^{(n)}}_1
\otimes \rho_2^{(n)},\eqno{(16)}$$
where $\rho_2^{(n)}$ is a
(second-subsystem) state and (16) is a
term in a final Hermitian Schmidt
canonical expansion of $\rho_{12}$.\\

Any biorthogonal decomposition of a
composite-system statistical operator
$\rho_{12}$ (into two or more terms)
can be continued in each term
separately into a Schmidt canonical
expansion of $\rho_{12}$ in terms of
Hermitian operators.

The biorthogonal decomposition is {\it
an intermediate step}. This is similar
to the case when we can partially
diagonalize the Hamiltonian of a
quantum system (due to some symmetry e.
g.). The diagonalization is then
continued separately with each
submatrix on the diagonal of the
Hamiltonian.

The continuation from a biorthogonal
mixture to a Hermitian Schmidt
canonical expansion can always be
performed, in principle, "by brute
force": diagonalizing the reduced
statistical superoperator $\hat \rho_1$
of the normalized supervector
$\ket{\rho_{12}}$ (analogously as it is
done for an ordinary state vector), and
by finding an invariant basis for
$V_1^{(a)}$ in each characteristic
subspace thus obtained \cite{Mes}.

The Hermitian Schmidt canonical
expansion of a composite-system
statistical operator will, hopefully,
find numerous applications in quantum
communication and information theory
because it lies at the basis of
entanglement. One of the applications
is evaluating all the twin observables.
This is illustrated elsewhere
\cite{Hor}.

\section{Weak twins and nonhermitian
 Schmidt canonical expansion}

For the sake of completeness it is
desirable to investigate decomposition
(3) also for a weak nontrivial twin
projector $P_1$. First, we take an
analytical view of Theorem 1 to realize
that the biorthogonality of the two
terms in (3) is connected with the twin
property (strong or weak), and the
strong twin property corresponds to the
hermiticity of the terms. Let us put
this more precisely.\\

\noindent {\it Definition 4.} A
decomposition $$
\rho_{12}=A_{12}+B_{12}$$ of a
composite-system statistical operator
$\rho_{12}$ into two linear operators
is {\it biorthogonal} if there exist
two opposite-subsystem projectors
$(P_1,P_2)$ such that $$
A_{12}=P_1A_{12}=P_2A_{12},\quad
0=P_1B_{12}=P_2B_{12};$$ $$
0=P_1^{\perp}A_{12}=P_2^{\perp}A_{12},\quad
B_{12}=P_1^{\perp}B_{12}=P_2^{\perp}B_{12}.
$$

It is clear from Theorem 1 that any
birthogonal mixture (of states) (6)
satisfies the generalized definition of
biorthogonality given in Definition 4.
Having in mind (3), it is also evident
that biorthogonality is equivalent to
the existence of a pair of twin
projectors (weak or strong). Finally,
the strongness property of the twins is
equivalent to the hermiticity of the
terms in (3), which results in having
statistical operator terms (and a
mixture).\\

\noindent {\it Theorem 3.} If
$(P_1,P_2)$ is a pair of {\it weak twin
projectors} for a composite-system
statistical operator $\rho_{12}$, then
the terms in (3) are super vectors, and
replacing each by a (nonhermitian)
 Schmidt canonical expansion, one obtains
 an expansion of the same kind for the
 entire statistical operator.\\

\noindent {\it Proof.} Since in $$
1\geq \Tr \rho_{12}^2=\Tr \rho_{12}
P_1\rho_{12}\enskip +\enskip \Tr
\rho_{12} P_1^{\perp}\rho_{12}$$ the
terms are nonnegative (as traces of
positive operators), the terms in (3)
are Hilbert-Schmidt operators, i. e.,
super vectors. Suppose we have expanded
the first term in (3) in the Schmidt
canonical way: $$
P_1\rho_{12}=c\sum_ir_i^{1/2}A_1^{(i)}
\otimes B_2^{(i)},$$ where $c$ is a
normalization constant (because the
statistical operator is not a super
state vector unless it is a pure
state). Since the LHS is invariant
under $P_1$, so is each first-subsystem
linear operator $A_1^{(i)}$, because
the second factors in the expansion
have unique corresponding first
factors. If we expand also the second
term in (3) in the Schmidt canonical
way
$$P_1^{\perp}\rho_{12}=c'\sum_jr_j'^{1/2}
C_1^{(j)} \otimes D_2^{(j)},$$ then,
analogously, invariance of each factor
$C_1^{(j)}$ under $P_1^{\perp}$
follows. This results in super vector
orthogonality: $$ \forall i,j:\qquad
\Tr \Big[(A_1^{(i)})^{\dagger}C_1^{(j)}
\Big]= \Tr \Big[
(A_1^{(i)})^{\dagger}P_1P_1^{\perp}
C_1^{(j)}\Big]=0.$$ The symmetrical
argument goes for the second factors
and $P_2$. Thus, replacing both terms
in (3) by their nonhermitian  Schmidt
canonical expansions, we have
biorthogonality between any term of the
first expansion and any term of the
second one. Therefore, we have an
expansion of the same kind for
$\rho_{12}$.\hfill $\Box$\\

It is now clear that also in the case
of weak twin projectors the
decomposition (3) can be {\it
continued}, but this time {\it to a
nonhermitian Schmidt canonical
expansion}.

As it was stated, I expect that
Hermitian Schmidt canonical expansion
of composite-system statistical
operators, and biorthogonal mixtures
that lead to it, will soon find
important application in quantum
communication and quantum information
theory. But, maybe, also the
nonhermitian version will be useful.

After all, a nonhermitian expansion
need not be wild and far fetched from
the physical point of view. Let me
illustrate this by the obvious fact
that a Schmidt canonical expansion of a
state vector $\ket{\Phi}_{12}$ $$
\ket{\Phi}_{12}=\sum_ir_i^{1/2}
\ket{i}_1\ket{i}_2,\qquad \forall
i\not= i':\quad \bra{i}_p
\ket{i'}_p=0,\enskip p=1,2$$
immediately results in a nonhermitian
 Schmidt canonical
expansion of the statistical operator
$\ket{\Phi}_{12}\bra{\Phi}_{12}$: $$
\ket{\Phi}_{12}\bra{\Phi}_{12}=
\sum_i\sum_{i'}r_i^{1/2}r_{i'}^{1/2}
\ket{i}_1\bra{i'}_1\otimes
\ket{i}_2\bra{i'}_2.$$

 Finally, let us return to separable
mixtures.

\section{Nontrivial twin projectors for
separable mixtures}

Let (9) be a general separable mixture.
Let us clarify under what conditions it
has nontrivial twin events.\\

\noindent {\it Theorem 4.} A general
separable mixture (9) has a nontrivial
twin projector $P_1$ {\it if and only
if} the set of all values of the index
"$k$" is the union of two
nonoverlapping subsets, say, consisting
of "$k'$" values and of "$k''$" values
respectively, and, when (9) is
rewritten accordingly: $$
\rho_{12}=\sum_{k'}w_{k'}\rho_1^{(k')}
\otimes \rho_2^{(k')}+\sum_{k''}w_{k''}
\rho_1^{(k'')}\otimes \rho_2^{(k'')},
\eqno{(17a)} $$ then one has
biorthogonality between the two groups
of terms: $$ \forall k',\enskip \forall
k'':\qquad
\rho_i^{(k')}\rho_i^{(k'')}=0,\quad
i=1,2.\eqno{(17b)}$$ Before we prove
the theorem, we first prove subsidiary
results.\\

\noindent {\it Lemma 1.} Let $$
\rho_{12}=\sum_mw_m\ket{\Psi^{(m)}}_{12}
\bra{\Psi^{(m)}}_{12}$$ be an arbitrary
pure-state mixture. Then, a pair of
subsystem observables $(A_1,A_2)$ are
twins for $\rho_{12}$ {\it if and only
if} they are twins for {\it all} pure
term-states.\\

\noindent {\it Proof. Necessity}
follows from the general result that
all twins of $\rho_{12}$ are also twins
of all state vectors from the
topological closure $\bar
\cR(\rho_{12})$ of the range of
$\rho_{12}$ (cf section 3, C1 in
\cite{HD00}). As well known, the
vectors
$\{\ket{\Psi^{(m)}}_{12}:\forall m\}$
span the mentioned subspace.

{\it Sufficiency} is obvious.\hfill
$\Box$\\

\noindent {\it Lemma 2.} Let $$
\rho_{12}=\sum_kw_k\rho_{12}^{(k)}$$ be
an arbitrary mixture. The pair
$(A_1,A_2)$ are twin observables for
$\rho_{12}$ {\it if and only if} they
are twin observables for {\it all} term
states $\rho_{12}^{(k)}$.\\

\noindent {\it Proof} is immediately
obtained from Lemma 1 if one rewrites
each term state as a pure-state
mixture.\hfill $\Box$\\

\noindent {\it Lemma 3.} An {\it
uncorrelated state} $\rho_1\otimes
\rho_2$ has only trivial twins.\\

\noindent {\it Proof} is an immediate
consequence of the fact that the tensor
factors of a nonzero uncorrelated
vector, say $a\otimes b$, are unique up
to an arbitrary nonzero complex number
$\alpha$, but if $a$ is replaced by
$\alpha a$, $b$ must be replaced by
$(1/\alpha )b$.

Applying this to supervectors  in case
of twins, we have $$ A_1\rho_1\otimes
\rho_2=\rho_1\otimes A_2\rho_2,$$ if
$A_1\rho_1=\alpha\rho_1$, then
$\rho_2=(1/\alpha)A_2\rho_2$.\hfill
$\Box$\\

\noindent {\it Proof of Theorem 3} now
immediately follows from Lemma 2 and
Lemma 3. Namely, the two groups of
terms stated in the Theorem, make up
the two terms in (3).\hfill $\Box$\\

\noindent {\it Corollary 3.} Nontrivial
twin events of a separable mixture (9)
are {\it necessarily strong twin
events}.\\

\noindent {\it Proof} is obvious if one
applies Lemmas 2 and  3 and if
adjoining is made use of.\hfill
$\Box$\\

\noindent {\it Corollary 4.} If
$(A_1,A_2)$ are nontrivial twin
observables for a separable mixture
(9), they are strong twin observables
(cf Definition 3), and the mixture
terms can be grouped into as many
biorthogonal groups of terms as there
are distinct characteristic values of
$A_1$ in $\cR(\rho_1)$ (generalization
of (17a,b)).\\

It is known that if a statistical
operator and a Hermitian operator
commute, then the corresponding state
can be written as a mixture so that
each term-state has a definite value of
the corresponding observable
\cite{Hro}. But, for the same
statistical operator, there are also
mixtures violating this.

To take an example, let us think of an
unpolarized mixture of spin-one-half
states: $\rho =(1/2)I$ (in the
two-dimensional spin factor space).
This statistical operator commutes with
$s_z$, nevertheless one can write down
the mixture $$ \rho
=(1/2)\Big(\ket{x,+}\bra{x,+}+
\ket{x,-}\bra{x,-}\Big)=(1/2)I,$$ in
which the term-states do not have a
definite value of the z-component.

It is interesting that in the case of a
separable mixture with a nontrivial
twin observable it is necessarily its
term-states that that have the sharp
detectable values of the corresponding
observable.\\

\noindent {\bf Acknowledgements}\\

\noindent The author is grateful to
Prof. Anton Zeilinger for his
invitation. Thanks are due also for the
financial support and hospitality of
the "Erwin Schr\"{o}dinger" Institute
in Vienna, where part of this work was
done. The author is indebted to
\v{C}aslav Brukner, who was of help by
asking the right question.

\end{document}